\documentclass[12pt]{article}
\usepackage{epsfig}

\begin{document}

\title{\Large \bf New Results on Nucleon Spin Structure}
\author{\large Jian-ping Chen \bigskip \\
{\it  Jefferson Lab, Newport News, Virginia, USA}}

\maketitle


\begin{center}
{\large {\bf Abstract}}\\
\end{center}
\par
\medskip
Recent precision spin structure data from Jefferson Lab have significantly 
advanced our knowledge of nucleon structure in the valence quark
(high-$x$) region and improved our understanding of
higher-twist effects, spin sum rules and 
quark-hadron duality.   
First, results of a precision measurement of the neutron spin 
asymmetry, $A_1^n$, in the high-$x$ region are discussed.
The new data shows clearly, for the first time, that $A_1^n$ becomes 
positive at high $x$.
They provide crucial input for the global fits to world data to extract 
polarized parton distribution functions. 
Preliminary results on $A_1^p$ and $A_1^d$ in the high-$x$
region have also become available. 
The up and down quark spin distributions in the nucleon were extracted. 
The results for $\Delta d/d$ disagree with the leading-order pQCD prediction assuming hadron helicity conservation.
Then, results of a precision measurement of 
the $g_2^n$ structure function to study higher-twist effects are presented.
The data show a clear deviation from the lead-twist contribution,
indicating a significant higher-twist (twist-3 or higher) effect.
The second moment of the spin structure functions and the twist-3
matrix element $d_2^n$ results were 
extracted at a high $Q^2$ of 5 GeV$^2$ from the measured
$A_2^n$ in the high-$x$ region in combination with existing world data and compared
with a Lattice QCD calculation.
Results for $d_2^n$ at low-to-intermediate $Q^2$ from 0.1 to 0.9 GeV$^2$ were also 
extracted from the JLab data. In the same $Q^2$ range, 
the $Q^2$ dependence of the moments of the nucleon spin structure functions 
was measured, providing a unique bridge linking the quark-gluon picture of 
the nucleon and the coherent hadronic picture. Sum rules and
generalized forward spin polarizabilities were extracted and compared with
Chiral Perturbation Theory calculations and phenomenological models. 
Finally, preliminary results on the resonance spin structure functions in the
$Q^2$ range from 1 to 4 GeV$^2$ were presented, which, in combination with DIS
data, will enable a detailed study of the quark-hadron duality in spin 
structure functions.

\section{Introduction and Motivation}
\par
Since the `spin crisis'\cite{spin}, substantial efforts, both theoretical and 
experimental, have been devoted to understanding the nucleon's spin puzzle. 
A new generation of experiments were carried out in the 1990s
at SLAC, CERN and DESY. These experiments concluded that the quarks carry 
about $ 20-30\%$ of the nucleon spin. The rest of the nucleon spin should come 
from the quark orbital angular momentum (OAM) and the gluon total angular 
momentum. 
The Bjorken sum rule\cite{Bjorken}, a fundamental sum rule of the spin 
structure function based on QCD, was verified to an accuracy of
better than $10\%$. Attempts have been made to extract the parton spin
distributions from a global analysis of the polarized 
deep-inelastic-scattering data. The uncertainties are much larger than those 
of the unpolarized parton distribution due to the fact that the polarized 
data coverage in $x$ adn $Q^2$ is much less extensive than that of the unpolarized data.   

Recently, the high polarized luminosity available at Jefferson 
Lab has allowed the study of the nucleon spin structure with 
an unprecedented precision, enabling us to access the
valence quark (high-$x$) region\cite{e99117} and also to expand the study to the
second spin structure function, $g_2$\cite{e97103}. Furthermore, the moments of the spin
structure functions\cite{chen05} were measured\cite{e94010} and the spin sum 
rules\cite{e94010,bjsum}, polarizabilities\cite{e94010}
and quark-hadron duality\cite{e01012,RSS} were studied.

\subsection{Inclusive Polarized Electron-Nucleon Scattering}
\par
For inclusive polarized electron scattering off a polarized 
nucleon target, 
the cross section depends on four structure functions, $F_1(Q^2,x)$, 
$F_2(Q^2,x)$, $g_1(Q^2,x)$ and $g_2(Q^2,x)$, where 
$F_1$ and $F_2$ are the unpolarized structure functions 
and $g_1$ and g$_2$ the polarized structure functions. 
In the quark-parton model, 
$F_1$ or $F_2$ gives the quark momentum 
distribution and $g_1$ gives the quark spin distribution.
Another physics quantity of interest is the virtual photon-nucleon 
asymmetry $A_1$
\begin{equation}
A_1={g_1-(Q^2/\nu^2) g_2 \over F_1} \approx {g_1 \over F_1}.
\end{equation}

\subsection{Spin structure in the valence quark (high-$x$) region}
\par
The high-$x$ region is of special interest, because this is where the valence 
quark contributions are expected to dominate.
With sea quark and explicit gluon contributions expected not to be
important, it is a clean region to test our understanding of nucleon
structure. Relativistic constituent quark models\cite{vqm}
should be applicable in this region
and perturbative QCD\cite{pQCD} can be used to make predictions in the large
$x$ limit. 

To first approximation, the constituent quarks in the nucleon are
described by SU(6) wavefunctions.
SU(6) symmetry leads to the following predictions\cite{SU6}: 

\begin{equation}
A_1^p=5/9;\ \ A_1^n=0; \ \ \Delta u/u=2/3; \ \ 
\Delta d/d=-1/3.
\label{eq:SU6}
\end{equation}

Relativistic Constituent Quark Models (RCQM) with broken SU(6) symmetry, e.g., 
the hyperfine 
interaction model\cite{vqm}, lead to a dominance of a `diquark' 
configuration 
with the diquark spin $S=0$ at high $x$. This implies that as $x\rightarrow1$:
\begin{equation}
 A_1^p\rightarrow 1;\ \
   A_1^n\rightarrow 1;\ \ \Delta u/u \rightarrow 1;\ \ {\rm and} \ \ 
\Delta d/d \rightarrow -1/3.
\label{eq:rnpqcd}
\end{equation}
\noindent In the RCQM, relativistic effects lead to 
non-zero quark orbital angular momentum and reduce the valence quark 
contributions to the nucleon spin from 1 to $0.6 - 0.75$.
 
Another approach is leading-order pQCD\cite{pQCD}, which assumes the 
quark orbital angular momentum to be negligible and leads to hadron helicity 
conservation. 
It yields:  

\begin{equation}
A_1^p\rightarrow 1;\ \
   A_1^n\rightarrow 1;\ \ 
\Delta u/u \rightarrow 1;\ \ {\rm and} \ \ 
\Delta d/d \rightarrow 1.
\label{eq:rnppqcd}
\end{equation}
\noindent
Not only are the limiting values as $x\rightarrow 1$ important, but also
the behavior in the high-$x$ region. How $A_1^n$ and  $A_1^p$ 
approach their limiting values when $x$ approaches 1, is sensitive to
the dynamics in the valence quark region. 

\subsection{The $g_2$ structure function and the $d_2$ moment}
\par
$g_2$, unlike $g_1$ and $F_1$, can not be
interpreted in the simple quark-parton model. To understand $g_2$ properly, 
it is best to start with the operator product expansion (OPE) method.
In the OPE, neglecting quark masses, $g_2$ can be cleanly separated into a
twist-2 and a higher twist term:
  \begin{eqnarray}g_2(x,Q^2) = g_2^{WW}(x,Q^2) +g_2^{H.T.}(x,Q^2)~.
  \end{eqnarray}
The leading-twist term can be determined from 
$g_1$ as\cite{WW}
  \begin{eqnarray}
   g_2^{WW}(x,Q^2) = -g_1(x,Q^2) + \int _{x}^1 \frac{g_1(y,Q^2)}{y} dy~,
  \end{eqnarray}
and the higher-twist term arises from quark-gluon correlations.
Therefore, $g_2$ provides a clean way to study higher-twist effects.
In addition, at high $Q^2$, the $x^2$-weighted moment, $d_2$, 
is a twist-3 matrix element and is related to the color 
polarizabilities\cite{d2}:
\begin{equation}
d_2 = \int _{0}^{1} x^2 [g_2(x)-g_2^{WW}(x)] dx.
\end{equation}
Predictions for $d_2$ exist from various models and lattice QCD.

\subsection{Moments and sum rules of spin structure functions}
\par
Sum rules involving the spin structure of the nucleon offer an important opportunity to study QCD. In recent years
the Bjorken sum rule at large $Q^2$ and 
the Gerasimov, Drell and Hearn (GDH) sum rule\cite{gdh} at $Q^2=0$
have attracted large experimental and theoretical\cite{dre01} efforts that have provided us with rich information. 
A generalized GDH sum rule\cite{ggdh} connects the GDH sum rule with the 
Bjorken sum 
rule and provides a clean way to test theories with experimental data over the
entire $Q^2$ range. 
Spin sum rules relate the moments of the spin structure functions to the 
nucleon's static properties (as in Bjorken or GDH sum rules) or
real or virtual Compton amplitudes, which can be calculated theoretically
(as in the generalized GDH sum rule or the forward spin polarizabilities). 
Refs. \cite{chen05,dre04} provide comprehensive reviews on this subject.

\subsection{Quark-hadron duality in spin structure functions}
\par
Quark-hadron duality was first observed  
for the spin-independent structure function $F_2$. In 1970, Bloom and Gilman\cite{BG} noted the nucleon 
resonance data averaged follows the DIS scaling curve. Recent precision 
data\cite{F2dual} confirm quark-hadron duality in the unpolarized proton
structure function. Efforts are ongoing to investigate quark-hadron duality
in polarized structure functions\cite{MEK}. It was predicted that in 
the high-x region at high enough $Q^2$, the resonances will have a similar 
behavior as the DIS. Results from HERMES\cite{HERMES} and CLAS\cite{eg1} 
show the proton spin structure function $g_1^p$ approaching
duality. The study of 
quark-hadron duality will aid in the study of the higher-twist effects and the 
high-x behavior in DIS. 

\section{Recent results from Jefferson Lab}
\par
The Thomas Jefferson National Accelerator Facility (Jefferson Lab, or JLab, 
formerly known as CEBAF - Continuous Electron Beam Accelerator Facility)
is located in Newport News, Virginia, USA. The accelerator produces 
a continuous-wave electron beam
of energy up to 6 GeV. An energy upgrade to 12 GeV is planned in the next
few years. The electron beam with a current of up to 180 $\mu$A is 
polarized up to $85\%$ by illuminating a 
strained GaAs cathode with polarized laser light. The electron beam is deflected 
to three experimental halls (Halls A, B and C) where electron beam can be
scattered
off various nuclear targets. The scattered electrons and knocked 
out particles are detected in the halls with various spectrometer detector 
packages. The experiments reported here are from inclusive electron scattering
where only the scattered electrons are detected.
The neutron results presented here are from
Hall A\cite{NIMA} where there are two High Resolution Spectrometers (HRS) 
with momentum up to 4 GeV/c. A polarized $^3$He target\cite{He3}, with in-beam polarization of about $40\%$,
provides an effective polarized neutron target. The polarized luminosity reached is $10^{36}$ s$^{-1}$cm$^{-2}$.
The detector package consists of vertical
drift chambers (for momentum analysis 
and vertex reconstruction), scintillation counters (data acquisition 
trigger) and \v{C}erenkov counters and lead-glass calorimeters 
(for particle identification (PID)). The $\pi^-$ were sorted from e$^-$ with an
efficiency better than 99.9\% .
Both HRS spectrometers were used to double the statistics and constrain the 
systematic uncertainties by comparing the cross sections extracted using each HRS.
The proton and deuteron results are from 
Hall B\cite{NIMB}, where there is the CEBAF Large Acceptance Spectrometer (CLAS) 
and Hall C\cite{HallC}, where there arr the High Momentum Spectrometer (HMS) and the Short Orbit Spectrometer (SOS). Polarized solid $NH_3$ and $ND_3$ 
targets\cite{NH3} using dynamical nuclear polarization were used. Polarizations
up to $70\%$ for $NH_3$ and up to $30\%$ for $ND_3$ were achieved.

  \subsection{Precision measurements of $A_1$ at high-$x$}
\par
In 2001, JLab experiment E99-117\cite{e99117} was carried out in Hall A
to measure $A_1^n$ with high precision in the $x$ region from 0.33 to 0.61
($Q^2$ from 2.7 to 4.8 GeV$^2$). 
Asymmetries from inclusive scattering of 
a highly polarized 5.7 GeV electron beam 
on a high pressure ($>10$ atm) (both longitudinally and
transversely) polarized $^3$He target were measured. 
Parallel and perpendicular asymmetries
were extracted for $^3$He. After taking into account the beam and target 
polarizations and the dilution factor,
they were combined to form $A_1^{^3He}$. Using the most recent 
model\cite{model}, nuclear
corrections were applied to extract $A_1^n$. The results on $A_1^n$
are shown in the left panel of Fig. 1. 

The experiment greatly improved the precision
of data in the high-$x$ region, providing the first evidence that 
$A_1^n$ becomes positive at large $x$, showing clear SU(6) symmetry 
breaking. The results are in good agreement with the LSS 2001 pQCD
fit to previous world data\cite{LSS2001} (solid curve) and 
the statistical model\cite{stat} (long-dashed curve).
The trend of the data is consistent with the RCQM\cite{vqm} predictions
(the shaded band). The data disagree with the predictions from the 
leading-order pQCD models\cite{pQCD} (short-dashed and dash-dotted curves).
These data provide crucial input for the global fits to the world data to 
extract the 
polarized parton distribution functions and the extractions of 
higher-twist effects. 

\begin{figure}[!ht]
\parbox[t]{0.5\textwidth}{\centering\includegraphics[bb=-20 -28 402 455, angle=0,width=0.5\textwidth]{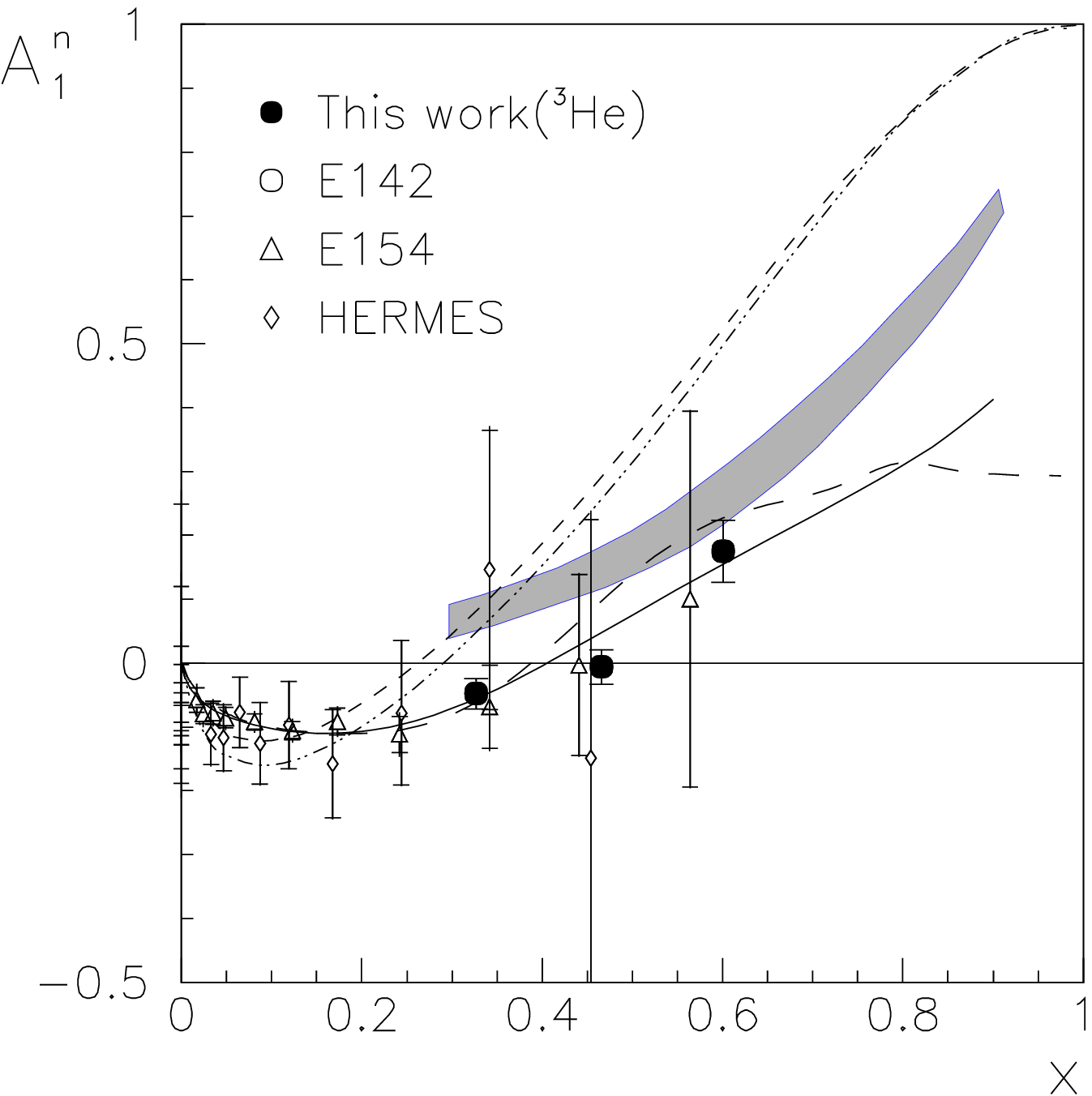}}
\parbox[t]{0.5\textwidth}{\centering\includegraphics[bb=-70 -28 352 455, angle=0,width=0.5\textwidth]{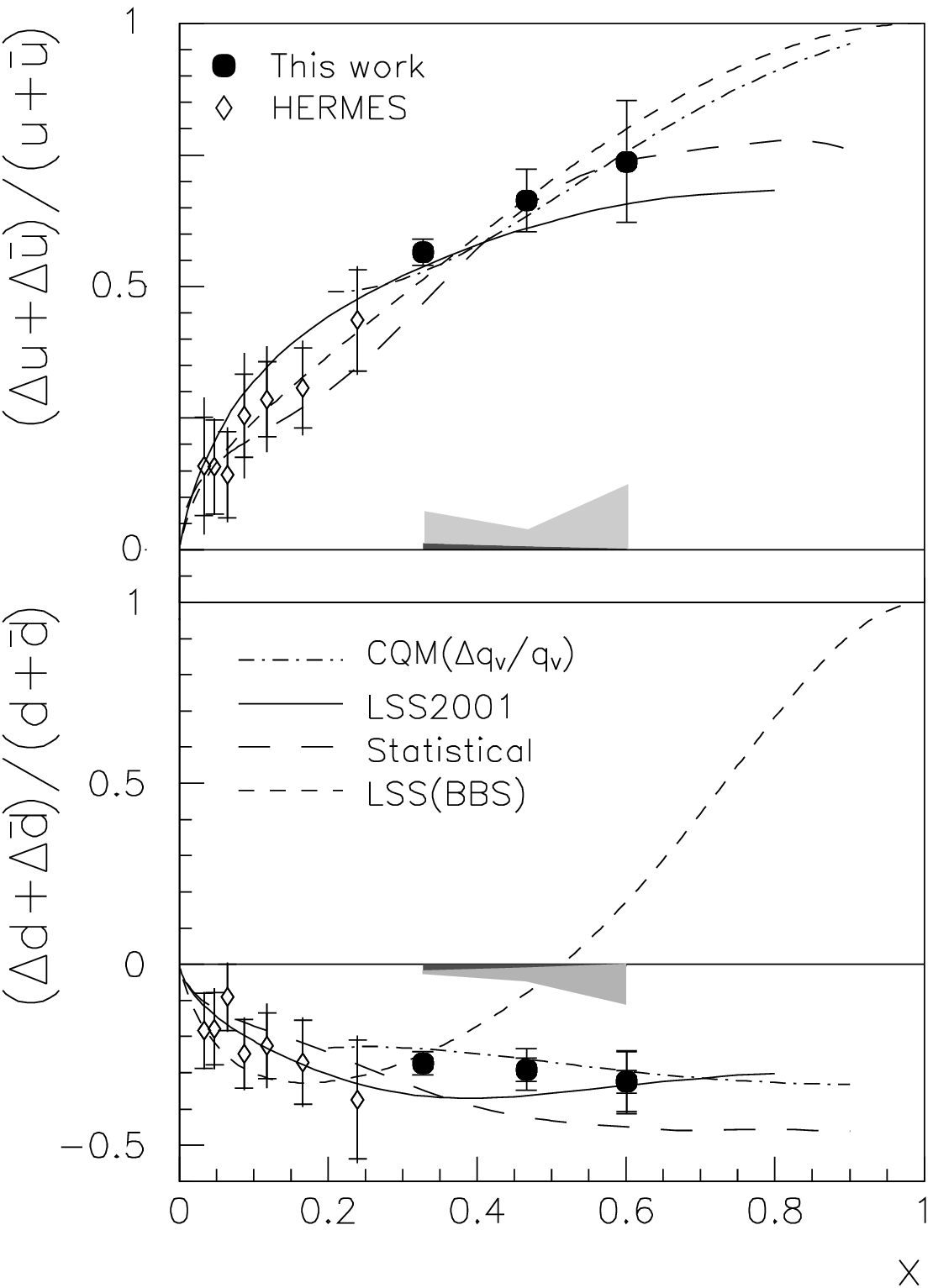}}
\caption {A$_1^n$, $\Delta u/u$ and $\Delta d/d$ results from JLab experiment E99-117, compared with 
the world data and theoretical predictions.}
\end{figure}
\medskip

In the leading-order approximation,
the polarized quark distribution functions $\Delta u/u$ and $\Delta d/d$ were 
extracted from our neutron data combined with the world proton data. 
The results are shown
in the right panel of Fig. 1, along with predictions from the RCQM (dot-dashed
curves), leading-order pQCD (short-dashed curves), the LSS 2001 fits 
(solid curves) and the statistical model (long-dashed curves).  The results 
agree well with RCQM predictions as well as the LSS 2001 fits and statistical 
model but $\Delta d/d$ is in significant disagreement with the 
predictions from
leading-order pQCD models assuming hadron helicity conservation. This
suggests that effects beyond leading-order pQCD, such as the quark orbital
angular momentum, may play an important role in this kinematic region.

\subsection{Measurements of $A_1^p$ and $A_1^d$ in the high x region}
\par
Preliminary results of $A_1^p$ and $A_1^d$ from the Hall B {\it eg1} experiment\cite{eg1} 
have recently become available. The data cover the $Q^2$ range of 1.4 to 4.5 
GeV$^2$ for $x$ 
from 0.2 to 0.6 with an invariant mass larger than 2 GeV. Data in the resonance
region are also available.
The precision of the data
improved significantly over that of the existing world data.

\subsection{Precision $g_2$ and $d_2$ measurements and higher twist effects}
\par
A precision measurement of g$_2^n$ from JLab Hall A E97-103\cite{e97103} covered 
five different Q$^2$ values from 0.58 to 1.36 GeV$^2$ at x $\approx 0.2$. 
Results for $g_2^n$ are given in the left panel of 
Fig. 2. The light-shaded area in the plot 
gives the leading-twist contribution, obtained by fitting world data\cite{BB} and
evolving to the $Q^2$ values of this experiment. The systematic errors are 
shown as the dark-shaded area near the horizontal axis.   

\noindent
\begin{figure}[!ht]
\parbox[t]{0.29\textwidth}{\centering\includegraphics[bb=660 22 1112 425, angle=-90,width=0.29\textwidth]{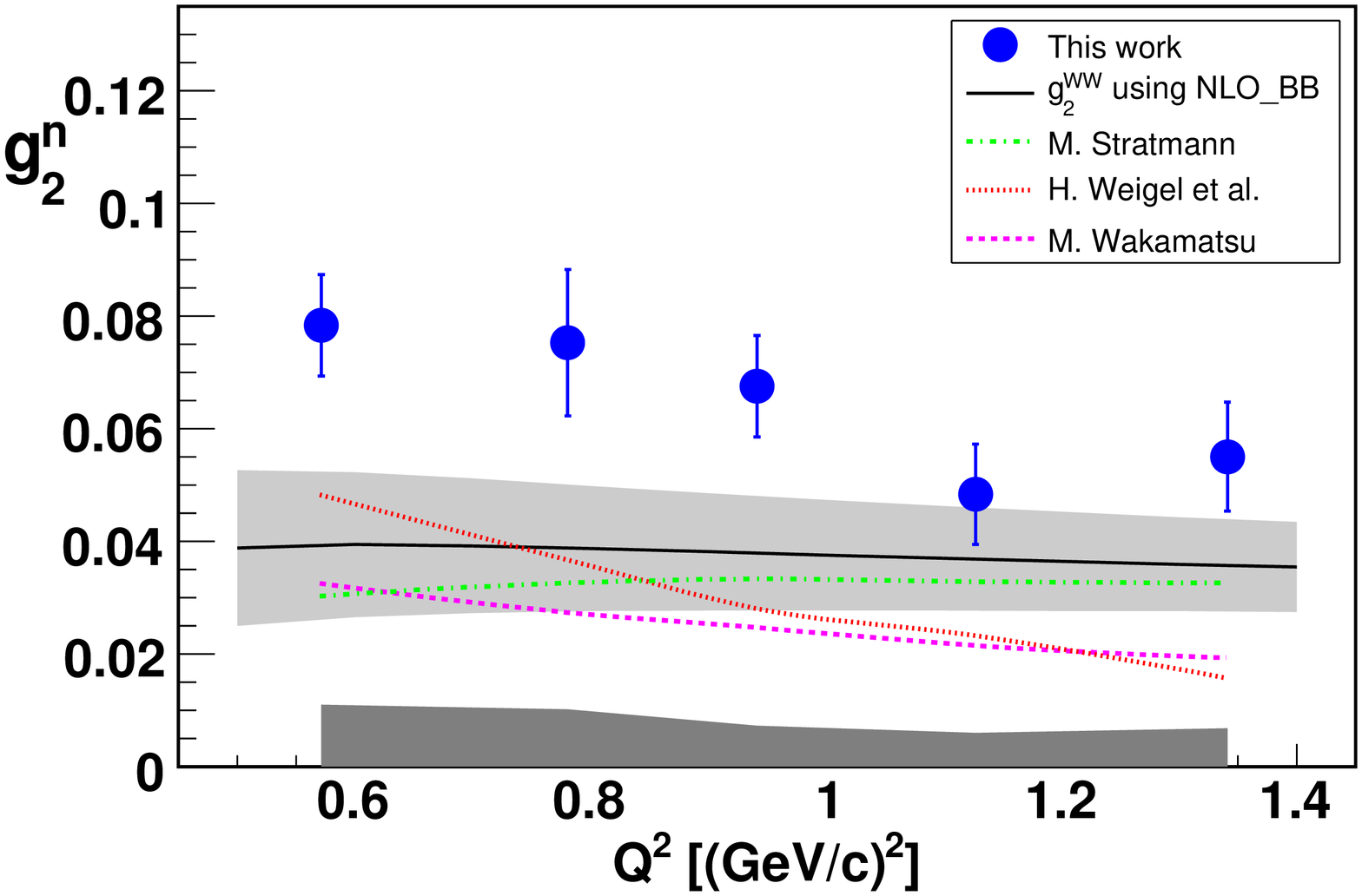}}
\parbox[t]{0.35\textwidth}{\centering\includegraphics[bb=-190 70 152 455, angle=0,width=0.35\textwidth]{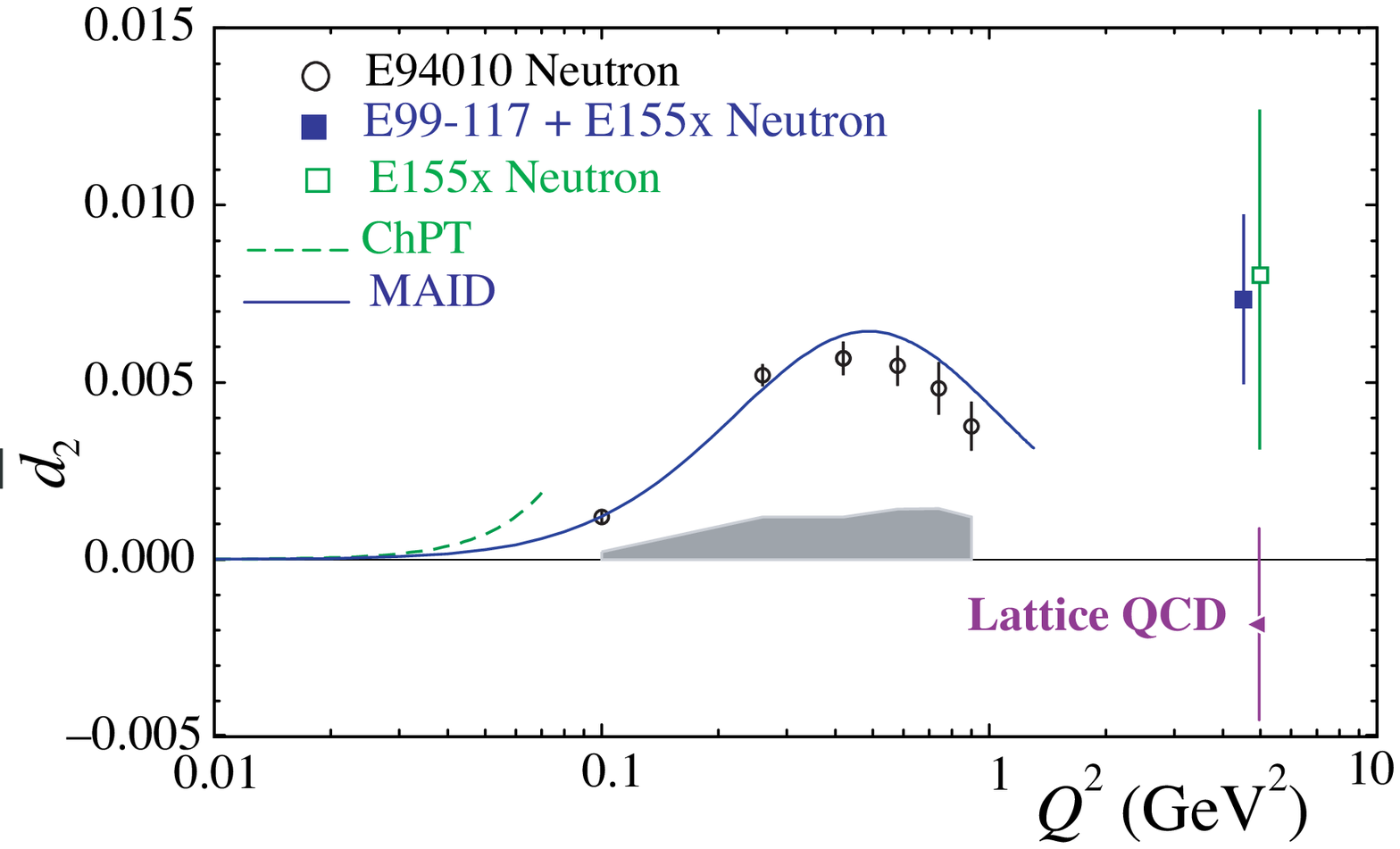}}
\vspace {-6cm}
\caption{Results for $g_2^n$ (left) and $d_2^n$ (right) from JLab Hall A.}
\end{figure}

The precision reached is more than an order
of magnitude improvement over that of the best world data\cite{E155x}.  The difference 
of g$_2$ from the leading twist part ($g_2^{WW}$)\cite{WW} is due to 
higher-twist effects and is sensitive to quark-gluon correlations. 
The measured g$_2^n$ values 
are consistently higher than g$_2^{WW}$.
For the first time, there is a clear indication that higher-twist effects 
become significantly positive at $Q^2$ below 1 GeV$^2$, 
while the bag model\cite{str} and Chiral Soliton model\cite{wei,wak} 
predictions of higher-twist effects are negative or close to zero. 
The $g_1^n$ data obtained from the same experiment agree with  
the leading-twist calculations within the uncertainties.

The second moment of the spin structure function $d_2$ is of special interest: 
at high $Q^2$, it is a twist-3 matrix element and can be calculated in 
lattice QCD. Experimentally, due to $x^2$ weighting, the contributions are dominated by the high-$x$ region and the problem of low-$x$ extrapolation is avoided. 
The Hall A experiment E99-117 also provide data on $A_2^n$ at high-$x$. 
The precision of the $A_2^n$ data is comparable to that of the 
best existing world data\cite{E155x} at high $x$. Combining these results with the world data, the second moment $d_2^n$ was extracted at an average $Q^2$ of 5
GeV$^2$.
Compared to the previously published result\cite{E155x}, the uncertainty on $d_2^n$ has 
been improved by about a factor of 2. The $d_2$ moment at high $Q^2$ 
has been calculated by Lattice QCD\cite{LQCD} and a number
of theoretical models. While a negative or near-zero value was 
predicted by Lattice QCD and most models, the new result for $d_2^n$ 
is positive. Also shown in Fig. 2 are the low $Q^2$ (0.1-1 GeV$^2$) results of the inelastic part of $d_2^n$ from another Hall A experiment E94-010\cite{e94010}, which were
compared with a Chiral Perturbation Theory calculation\cite{chpt} and a model prediction\cite{maid}. Detail of the experiment E94-010 and comparison with ChPT and model will be discussed below.

\subsection{Precision measurements of Moments and sum rules of spin structure functions}
\par
JLab E94-010\cite{e94010} in Hall A measured
the generalized GDH sum and the moments of the neutron spin structure functions
$\Gamma_1$ and $\Gamma_2$
in the low to intermediate $Q^2$ range. 
The measurement of doubly-polarized inclusive cross sections was performed
at five beam energies from 0.86 to 5.1 GeV 
at a scattering angle of $15.5^\circ$. 
Parallel and perpendicular cross-section differences were obtained,
from which $g_1$, $g_2$, 
$\sigma_{TT}$ and $\sigma_{LT}$ 
for $^3$He were extracted. Interpolation to constant $Q^2$ values was 
performed and
the GDH integrals were formed from pion threshold to $W^2=4$ GeV$^2$. 
Finally, nuclear corrections\cite{ciofi} were applied
to extract the GDH integral for the neutron.
The results are shown in the left-top panel of Fig. 3. 
 The higher energy contributions (for $W^2$ from
4 to 1000 GeV$^2$) were estimated using the parameterization of Thomas and 
Bianchi\cite{TB}. 

These data show a smooth but dramatic change in the value of the generalized 
GDH sum
from what was observed at high $Q^2$. While not unexpected from 
phenomenological models, these data illustrate the sensitivity to the transition
from partonic to hadronic behavior. The measured values of the first moment
of 
$g_1^n$ are shown in the left-middle panel of Fig. 3, along with the world data from SLAC and HERMES. Also shown are Chiral Perturbation Theory calculations and several model predictions. 
These data provide a precision data 
base for a twist expansion analysis at the higher end of the $Q^2$ range, a check for Chiral Perturbation Theory (ChPT)
calculations\cite{chpt} at the low end of the $Q^2$ range, and establish an important benchmark against which one can 
compare future calculations (such as Lattice Gauge Theory calculations).
The measured values of the first moment of $g_2^n$ are shown in the left-bottom 
panel of Fig. 3.
These results indicate the first validation of the Burkhardt-Cottingham sum 
rule\cite{BC}, $\Gamma_2=0$.
Also shown in the right-bottom panel is the Bjorken sum\cite{bjsum}, the
first moment of $g_1^p-g_1^n$, which were extracted from the E94-010 neutron data and the CLAS proton and deuteron results. 
\noindent
\begin{figure}
\parbox[t]{0.6\textwidth}{\centering\includegraphics[bb=104 172 554 655, angle=-90,width=0.4\textwidth]{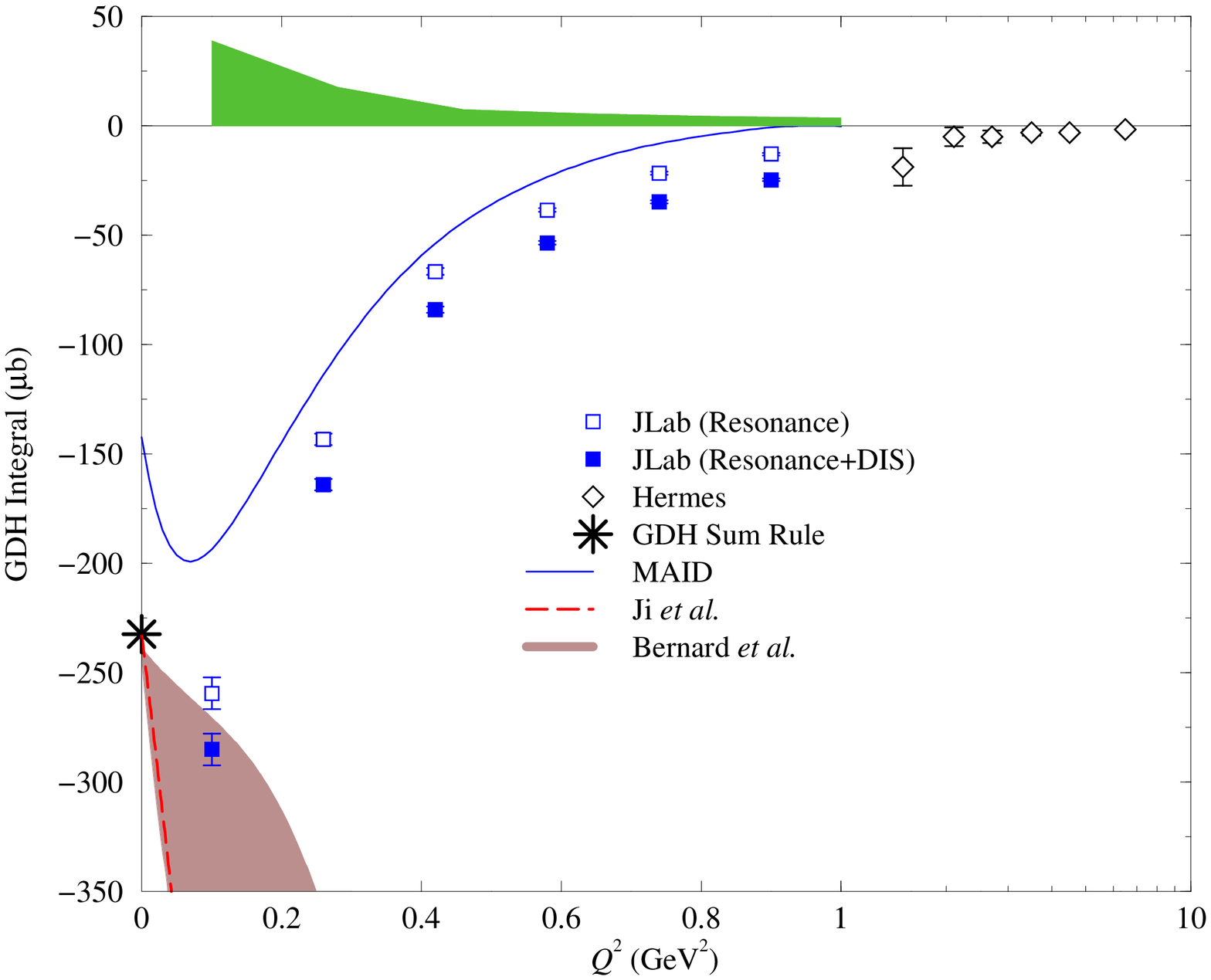}}
\parbox[t]{0.6\textwidth}{\centering\includegraphics[bb=104 272 554 755, angle=-90,width=0.4\textwidth]{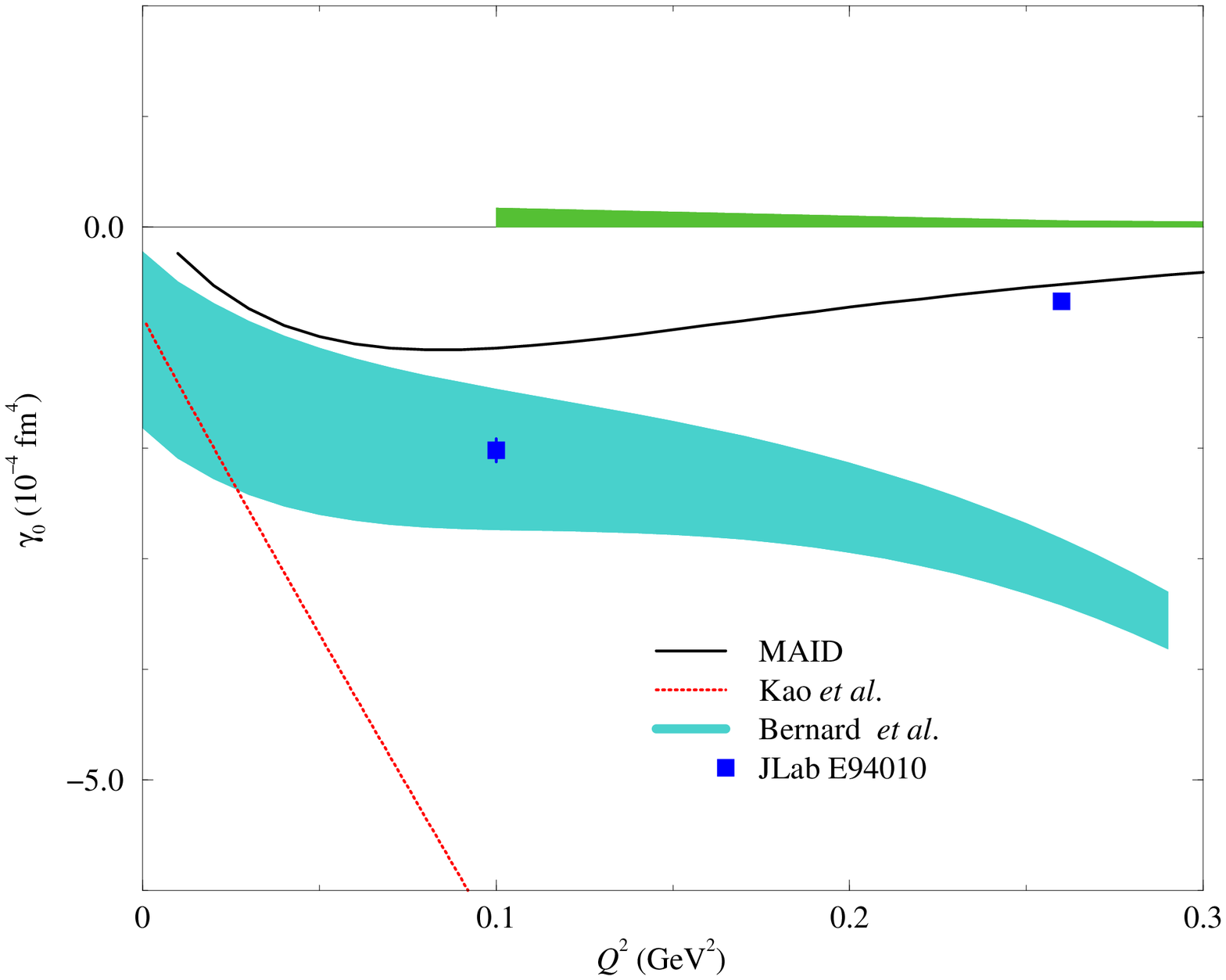}}
\parbox[t]{0.6\textwidth}{\centering\includegraphics[bb=104 172 556 655, angle=-90,width=0.4\textwidth]{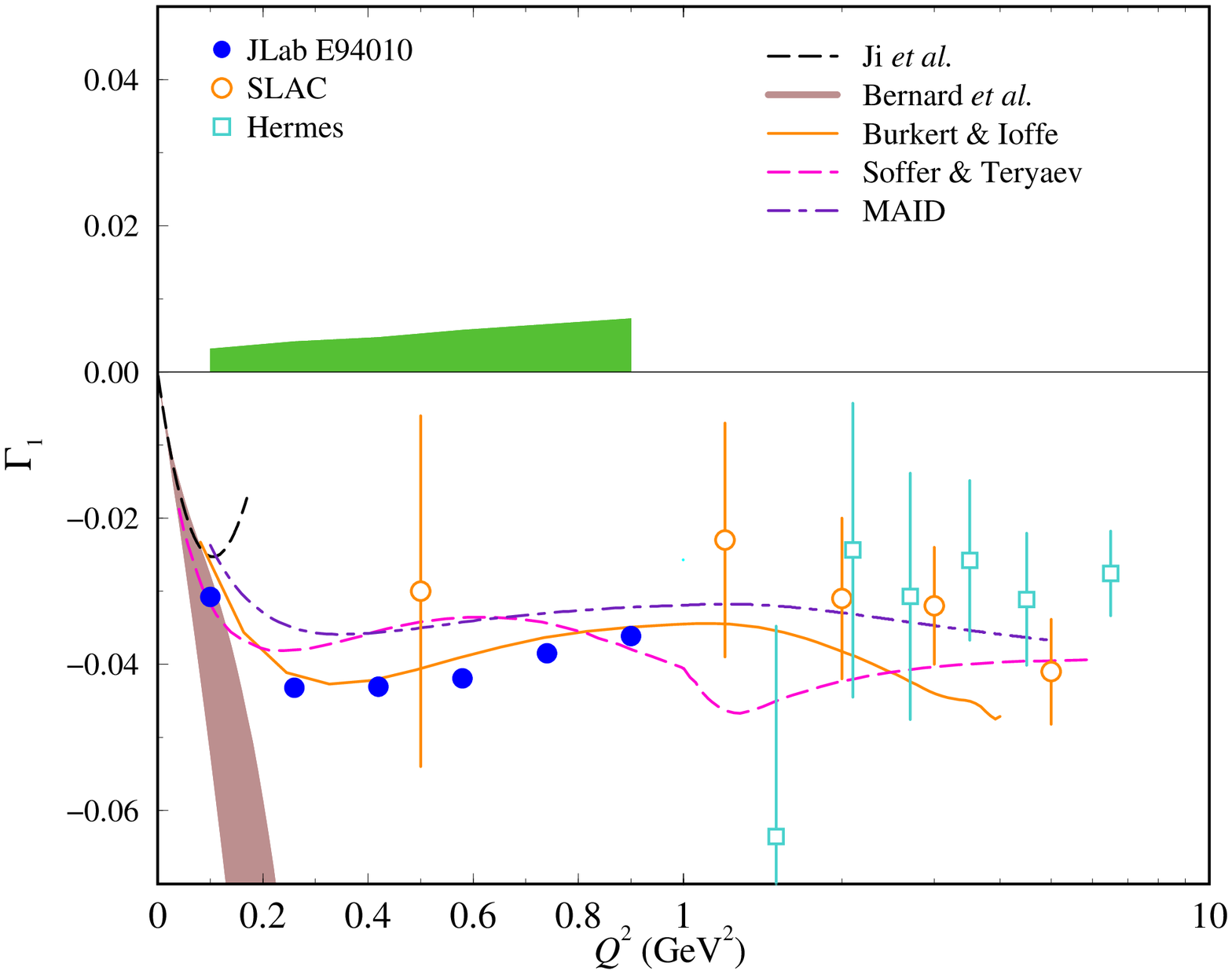}}
\parbox[t]{0.6\textwidth}{\centering\includegraphics[bb=104 272 554 755, angle=-90,width=0.4\textwidth]{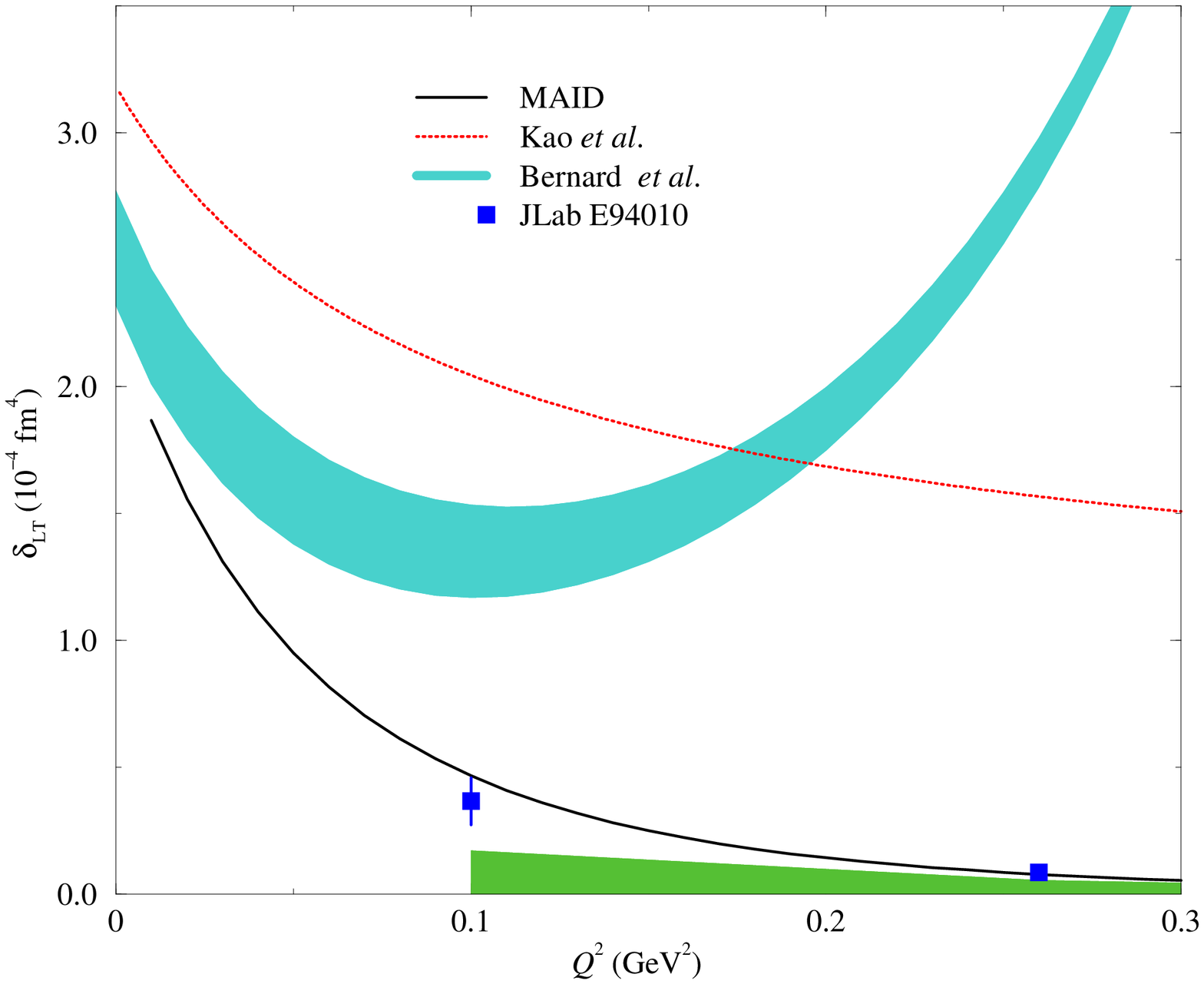}}
\parbox[t]{0.6\textwidth}{\centering\includegraphics[bb=104 172 556 655, angle=-90,width=0.4\textwidth]{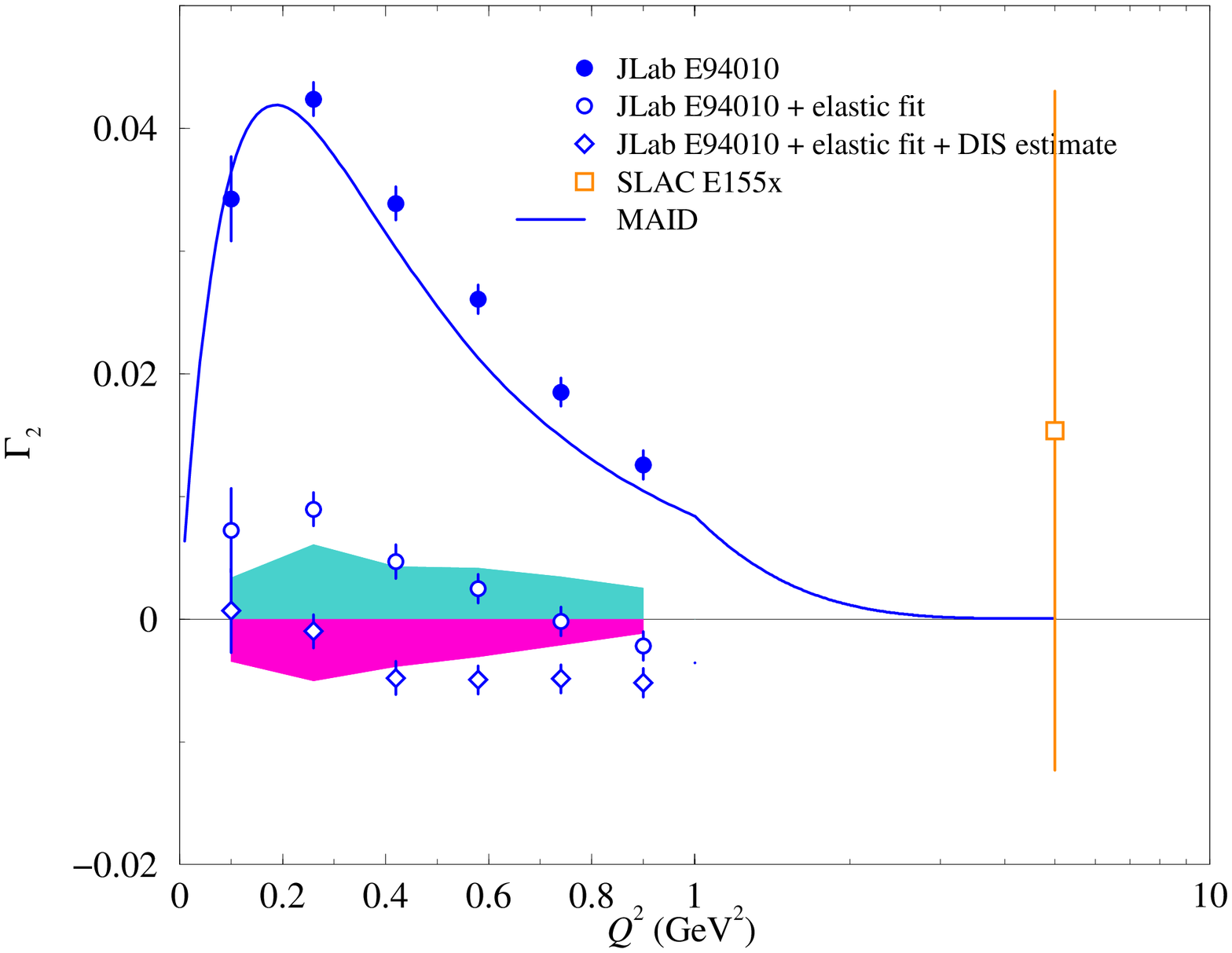}}
\parbox[t]{0.6\textwidth}{\centering\includegraphics[bb=202 504 675 104, angle=0,width=0.4\textwidth]{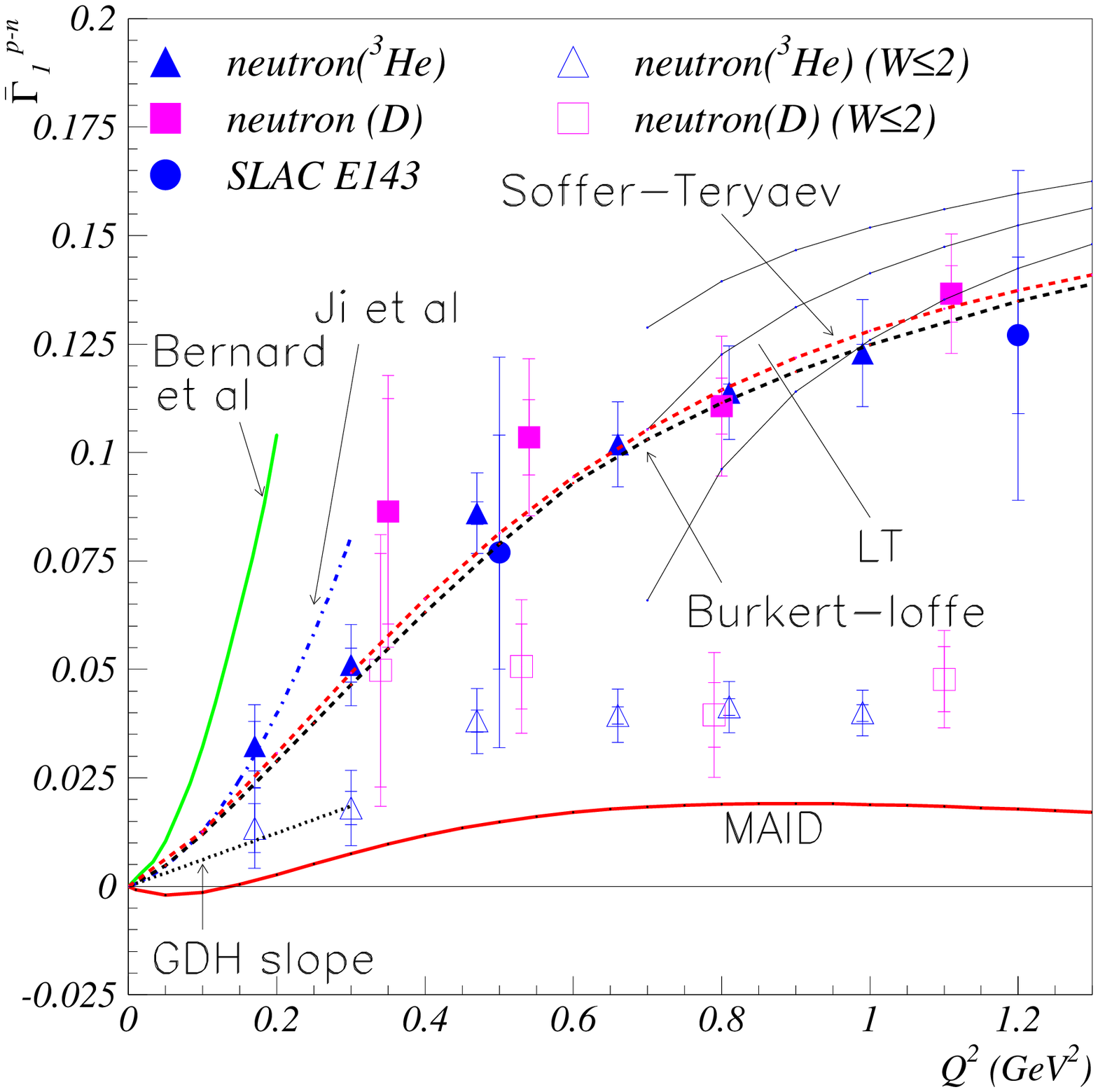}}
\vspace {1cm}
\caption {Comparisons of JLab E94-010 neutron results and the Bjorken sum (proton-neutron) results, which are extracted from E94-010 and CLAS data,
with world data, 
ChPT calculations and model calculations. The bands above the zero axes are the 
systematic uncertainties. The band below the zero axis on left-bottom panel shows the estimated uncertainties from low-$x$ extrapolation.}
\end{figure}

Higher ($x^2$ weighted) moments of the spin structure functions are related to 
the generalized forward spin polarizabilities $\gamma_0$ and
$\delta_{LT}$\cite{e94010}.
The right panels of Fig. 3 show the E94-010 results on $\gamma_0$ and $\delta_{LT}$, and the comparison with
the ChPT calculations and MAID model\cite{maid} predictions.
The relativistic baryon ChPT with resonance shows a good agreement with the 
data for $\gamma_0$ at $Q^2 = 0.1$ GeV$^2$. However, ChPT calculations deviate 
significantly from 
the data for $\delta_{LT}$, which was expected to be an excellent candidate
to check Chiral dynamics of QCD since it was not sensitive to the
dominating resonance ($\Delta$) contributions. This disagreement presents a 
real challenge to theorists.  

New experiments\cite{e97110,e03006} will extend the generalized GDH
sum measurements to very low $Q^2$ (down to $Q^2=0.01$ GeV$^2$),
below the turn-around point predicted by calculations
(at $Q^2 \approx 0.1$ GeV$^2$). 
ChPT calculations will be extensively tested at low $Q^2$ where they
are expected to be applicable.
Extrapolation to the real photon point
provides an alternative way to test the original GDH sum rule.
Data taking was completed for the neutron in Hall A in the summer of 2003. Analysis is underway. Data taking is planned in 2006 for the proton 
experiment in Hall B.

\subsection{Preliminary results of spin structure functions for Quark-hadron duality study}
\par
JLab E01-012\cite{e01012} ran successfully in early 2003 in Hall A. Asymmetries and cross sections were measured in the resonance region, for $Q^2$ range from 1 to 3.6 GeV$^2$, for inclusive scattering of polarized electrons on a
longitudinally and transversely polarized $^3$He target. The spin structure function $g_1$ and virtual photon asymmetry $A_1$ were extracted. 
Work is ongoing to extract the neutron results. These results, combined with the DIS results\cite{e99117}, will provide a test of quark-hadron duality in
the neutron spin structure functions. 


Preliminary results have also become available from the JLab Hall C experiment E01-006
\cite{RSS} on the 
proton spin structure in the resonance region. These data, combined with the world DIS data, will help study quark-hadron duality in the proton spin 
structure function.

\section{summary}
\par
In summary, the high polarized luminosity available at
JLab has provided us with high-precision data to study the nucleon
spin structure in a wide kinematic range. They shed
light on the valence quark structure and helped to 
understand quark-gluon correlations and study the transition 
between perturbative and non-perturbative regions of QCD.
\medskip

\footnotesize{
The work presented was supported in part 
by the U. S. Department of Energy (DOE)
contract DE-AC05-84ER40150 Modification NO. M175,
under which the
Southeastern Universities Research Association operates the 
Thomas Jefferson National Accelerator Facility.
}



\begin{thebibliography}{99}
\bibitem{spin} see, for example, B. W. Filippone and X. Ji, 
Adv. Nucl. Phys. {\bf 26}, 1 (2001)
\bibitem{Bjorken} J. D. Bjorken, Phys. Rev. {\bf 148}, 1467 (1966);
Phys. Rev. D{\bf 1}, 465 (1970)
\bibitem{e99117} X. Zheng, {\it et al.}, Phys. Rev. Lett. {\bf 92},
012004 (2004); X. Zheng, {\it et al.}, Phys. Rev. C {\bf 70}, 065207 (2004).
\bibitem{e97103} K. Kramer, {\it et al.}, Phys. Rev. Lett.{\bf 95},142002 (2005). 
\bibitem{chen05} J. P. Chen, A. Deur and Z. E. Meziani, to appear in Mod. Phys. Lett. A (2005); nucl-ex/0509007.
\bibitem{e94010} M. Amarian, {\it et al.}, Phys. Rev. Lett. {\bf 89},
242301 (2002); 
{\it ibid.}, {\bf 92}, 022301 (2004);
{\it ibid.}, {\bf 93}, 152301 (2004); Z. E. Meziani {\it et al.}, Phys. Lett. B {\bf 613}, 148 (2005).
\bibitem{bjsum} A. Deur, {\it et al.}, Phys. Rev. Lett. {\bf 93}, 212002 (2004).
\bibitem{e01012} JLab E01-012, Spokespersons, J. P. Chen, S. Choi and N. Liyanage.
\bibitem{RSS} JLab E01-006, Spokespersons, M. Jones and O. Rondon. 
\bibitem{vqm} N. Isgur, Phys. Rev. D {\bf 59}, 034013 (1999).
\bibitem{pQCD}
S. Brodsky, M Burkhardt and I. Schmidt, Nucl. Phys. 
B{\bf 441}, 197 (1995).
\bibitem{SU6} F. Close, Nucl. Phys. B {\bf 80}, 269 (1974).
\bibitem{WW}
S. Wandzura and F. Wilczek, Phys. Lett. B 72 (1977).
\bibitem{d2}
X. Ji and W. Melnitchouk, Phys. Rev. D {\bf 56}, 1 (1997).
\bibitem{gdh} S. B. Gerasimov, Sov. J. Nucl. Phys. {\bf 2}, 598 (1965);
S. D. Drell and A. C. Hearn, Phys. Rev. Lett. {\bf 162}, 1520 (1966).
\bibitem{dre01} D. Drechsel, S.S. Kamalov, and L. Tiator, Phys. Rev. D {\bf 63}, 114010 (2001). 
\bibitem{ggdh} X. Ji and J. Osborne, J. Phys. G {\bf 27}, 127 (2001).
\bibitem{dre04} D. Drechsel and L. Tiator, Ann. Rev. Nucl. Part. Sci.
{\bf 54}, 69 (2004).
\bibitem{BG} E. D. Bloom and F. J. Gilman, Phys. Rev. Lett. {\bf 25}, 1140 (1970).
\bibitem{F2dual} I. Niculescu, {\it et al.}, Phys. Rev. Lett. {\bf 85}, 1182 (2000); {\bf 85}, 1186 (2000).
\bibitem{MEK} W. Melnitchouk, R. Ent and C. Keppel, Phys. Rept. {\bf 406}, 127 (2005).
\bibitem{HERMES} A. Airapetian, {\it et al.}, Phys. Rev. Lett. {\bf 90}, 092002 (2003).
\bibitem{eg1dual} T. A. Forest, Proceedings of GDH2004, editors, S. Kuhn and J. P. Chen, World Scientific.
\bibitem{NIMA}Hall A collaboration: J. Alcorn \emph{et al.}, 
Nucl. Inst. Meth. A \textbf{522}, 294 (2004).
\bibitem{He3}J.S. Jensen, Ph.D. Thesis, California Institute of Technology, 2000; I. Kominis, Ph.D. Thesis,
Princeton University, 2000; /www.jlab.org/e94010/.
\bibitem{NIMB}
CLAS collaboration: B. A. Mecking \emph{et al.},
Nucl. Inst. Meth. A \textbf{503}, 513 (2003).
\bibitem{HallC}http://www.jlab.org/Hall\-C/
\bibitem{NH3} 
C. D. Keith \emph{et al}., Nucl. Inst. Meth. A \textbf{501}, 327 (2003).
\bibitem{model} F. Bissey, {\it et al.}, Phys. Rev. C {\bf 65}, 064317 (2002).
\bibitem{LSS2001} E. Leader, A. V. Sidorov and D. B. Stamenov, Eur. Phys. J.
C {\bf 23}, 479 (2002).
\bibitem{stat} C. Bourrely, J. Soffer and F. Buccella, Eur. Phys. J. 
C {\bf 23}, 487 (2002).
\bibitem{eg1} S. Kuhn, private communication and J. P. Chen, Proceedings of DIS05, AIP {\bf 792}, 961 (2005).
\bibitem{E155x} K. Abe, {\it et al.}, E155 collaboration, Phys. Lett.
B {\bf 493}, 19 (2000).
\bibitem{BB} J. Bl{\"{u}}mlein and H. Bottcher, Nucl. Phys. B {\bf 636}, 225 (2002).
\bibitem{str} M. Stratmann, Z. Phys. C {\bf 60}, 763 (1993).
\bibitem{wei} H. Weigel, Pramana {\bf 61}, 921 (2003).
\bibitem{wak} M. Wakamatsu, Phys. Lett. B {\bf 487}, 118 (2000).
\bibitem{LQCD} M. Gockeler {\it et al.}, Phys. Rev. D {\bf 63}, 074506 (2001).
\bibitem{chpt}X. Ji, C. Kao, and J. Osborne, Phys. Lett. B {\bf 472}, 1 (2000);
C.~W.~Kao, T. Spitzenberg and M. Vanderhaeghen, Phys. Rev. D {\bf 67}, 016001 (2003);
V. Bernard, T. Hemmert and Ulf-G. Meissner, Phys. Rev. D {\bf 67}, 076008
\bibitem{maid} D. Drechsel, S. Kamalov and L. Tiator, Phys. Rev.  D {\bf 63}, 114010 (2001)
\bibitem{ciofi} C. Ciofi degli Atti and S. Scopetta, Nucl. Phys. B{\bf 404}, 
223 (1997)
\bibitem{TB} E. Thomas and N. Bianchi, Nucl. Phys. B{\bf 82} (Proc. Suppl.), 
256 (2000)
\bibitem{BC} H. Burkhardt and W. N. Cottingham, Ann. Phys. {\bf 56}, 453 (1970)
\bibitem{e97110}JLab E97-110, Spokespersons, J. P. Chen, A. Deur and 
F. Garibaldi.
\bibitem{e03006} JLab E03-006, Spokespersons, M. Battaglieri, A. Deur, R. De Vita and M. Ripani.  
\end{thebibliography}
\end{document}